\documentclass[aps,prb,twocolumn,groupedaddress,showpacs,floatfix]{revtex4}
\usepackage{epsfig}
\usepackage{graphicx,floatflt}
\usepackage{epsf} 
\begin{document}
%

\title{Ab-initio calculations of the optical properties of the Si(113)
  3$\times$2ADI surface}

\author{K. Ga\'al-Nagy}
\email{katalin.gaal-nagy@physik.uni-r.de}
\author{G. Onida}
\affiliation{European Theoretical Spectroscopy Facility (ETSF),
  CNISM-CNR-INFM, and Dipartimento di Fisica dell'Universit\`a degli
  Studi di Milano, via Celoria 16, I-20133 Milano, Italy}

\date{\today}
%
\begin{abstract}
We investigated the stable silicon (113) surface with a 3$\times$2ADI
reconstruction by ab-initio methods. The ground state properties have
been obtained using the density-functional theory. We present the
dispersion of the electronic band structure, where the surface bands
have been distinguished from the projected bulk bands by calculating
their localization in the slab. The optical spectra, here the
reflectance anisotropy (RAS), have been obtained within the
independent particle random phase approximation. We identified surface
features in the spectra tracing them back to the responsible
electronic states and, studied their localization in the slab. A
comparison with available experimental data for the band structure and
the RAS shows a good agreement.
\end{abstract}
\maketitle

%
%
\section{Introduction}\label{Intro}
Although being of high index, the vicinal Si(113) surface is one of
the stable silicon (Si) surfaces.\cite{Eag1993b} This surface is of
technological interest, since atomically smooth, ultrathin oxide films
can be grown on it.\cite{Mus2001a} Hence, it is dealt as a candidate
for the next generation of wafers,\cite{Mus2001b} and has potential
applications in nanostructures technology. Si(113) can be used as a
substrate for the self-assembled growth of germanium (Ge)
nanodots\cite{Zha2004c} and nanowires\cite{Hal2002, Zha2002c,
Sum2002}, as well as Ge\cite{Kna1992, Omi1999} and SiGe\cite{Han2005}
islands.  Besides the technological interest, Si(113) is also of
fundamental one, because it shows phase transitions between its
3$\times$2 and 3$\times$1 reconstructions. These phase transitions can
be induced by temperature \cite{Yan1990a, Xin1990, Hib1997, Suz1998,
Hwa2001, Hwa2002} and by contamination.\cite{Jac1993, Myl1989,
Yan1990b, Sch1995b, Hwa2000, Kim2001, Zha2005a, Zha2005b} The latter
ones affect also the discussion about the surface reconstruction. In
spite of the measurement of a 3$\times$1 reconstruction,\cite{Had1993}
most of the experiments show at room temperature (RT) a 3$\times$2
surface periodicity.\cite{Myl1989, Kna1991, Jac1993, Sch1995b,
Sak1996, Hib1997, Cha1999b, Hwa1999a, Zha2004c} The finding of the
3$\times$1 periodicity might be due to contamination\cite{Jac1993} and
the 3$\times$2 one is assumed as the surface unit of clean Si(113) at
RT.

\begin{figure}[b!]
\begin{minipage}{8.6cm} 
  \epsfig{figure=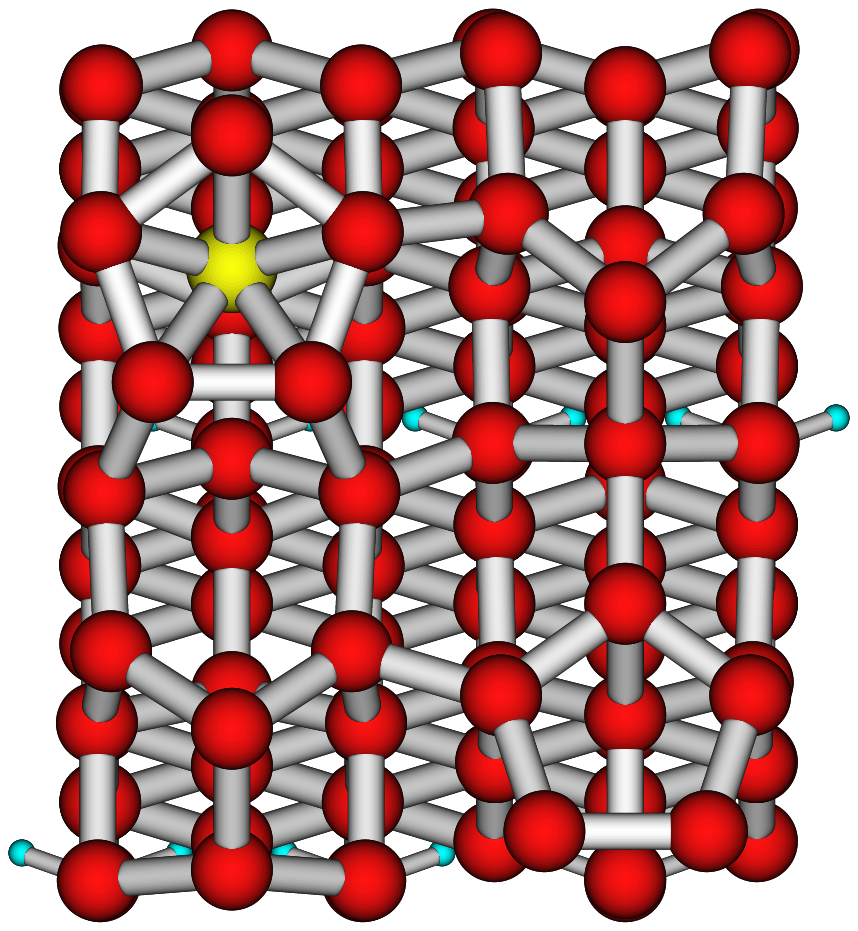,width=4.3cm,angle=0}
  \epsfig{figure=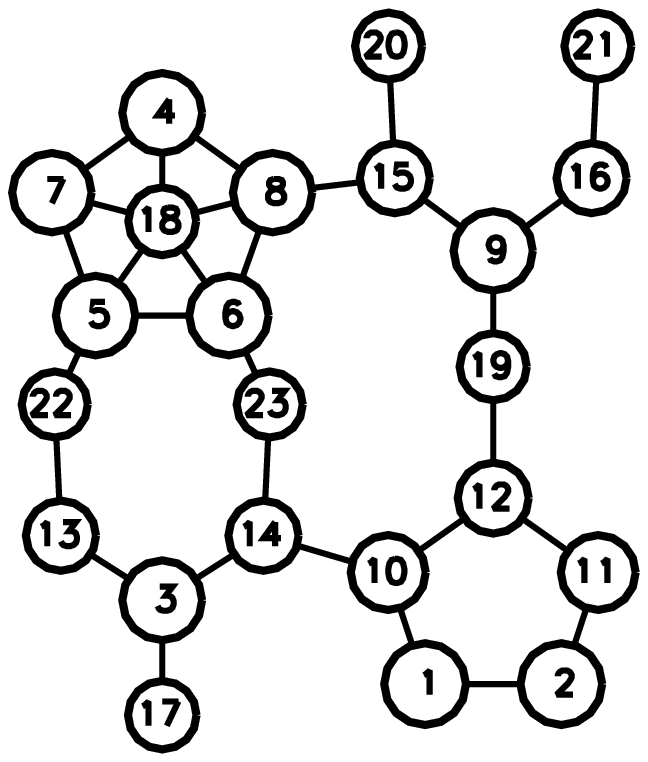,width=3.8cm,angle=0}
\end{minipage}
  \caption{(color online) Top view of the Si(113)3$\times$2ADI cell
  ($x$, $y$ plane corresponding to $[1\bar{1}0]$ and $[33\bar{2}]$,
  respectively). Silicon atoms are in large dark spheres (red). The
  interstitial Si atom (atom 18) is depicted as a large light
  (yellow) sphere. Hydrogen atoms are small light spheres (blue). For
  a side view see Figs.~\ref{Pic_LayerByLayer} and
  \ref{Pic_chargeTamm}. Surface atoms (see sketch) include two
  pentagons with dimers (atom pairs 1-2 and 5-6) and two adatoms
  (atoms 3 and 9).
  }\label{Pic_Structure}
\end{figure}

Various surface reconstructions have been proposed for the
Si(113)3$\times$2 surface. The most probable surface model is the ADI
(adatom-dimer-interstitial) reconstruction of
D\c{a}browski,\cite{Dab1994} which is in agreement with experimental
high-resolution transmission electron microscopy (HRTEM)\cite{Tak2001,
Tak2000} and scanning tunneling microscopy (STM)\cite{Kna1991,
Dab1994,Dab1995c, Sak1996, Hib1997, Zha2004c} images. It has been
confirmed by other theoretical investigations, too.\cite{Dab1995c,
Dab1995d, Ste2003b, Lee2004} A top view of the surface unit cell is
shown in Fig.~\ref{Pic_Structure}.  The surface reconstruction shows
two adatoms, and two pentagons with a dimer along the $[1\bar{1}0]$
direction, with one pentagon hosting an interstitial atom in its
center.

Earlier structured models were based on the structure proposed by
Ranke,\cite{Ran1990} Si(113)3$\times$1AD, which is similar to the ADI
one. It can be deduced from the ADI model by eliminating the
interstitial atom and relaxing the atoms in $z$ direction. Despite
being intrinsically 3$\times$1, also in the 3$\times$1AD model the
3$\times$2 periodicity can be obtained either by canceling one
pentagon in each 3$\times$2 surface unit cell
(``surface-void''[\onlinecite{Ran1990}]), or by manipulating the $z$
coordinates of the pentagons in order to break the symmetry
(``corrugated''[\onlinecite{Fen1997}],
``puckered''[\onlinecite{Cha1999b, Wan1996}],
``shifted''[\onlinecite{Kim2003}], etc.). Since the 3$\times$1
periodicity is also interesting for the phase transition, being the
periodicity of the high-temperature phase, besides the 3$\times$1AD
model also a 3$\times$1AI one, with an interstitial atom at each
pentagon has been proposed.\cite{Dab1995c, Lee2003b} However, also
other surface models have been assumed, generally based on the models
of Ranke or D\c{a}browski.\cite{Ike1999, Suz1999}

From the theoretical point of view, the most stable surface (i.e., the
one with the lowest surface energy) should be selected. For some of
the models above, the surface energy has been calculated,
unfortunately by using different methods. There exist some studies
assuming more than one surface reconstruction\cite{Ste2003a, Lu2005,
Lee2004, Dab1994, Dab1995c} within the same framework, but none of
them considered all available surface models for a comparison. Most of
the calculations have shown that the Si(113)3$\times$2ADI surface has
the lowest surface energy with respect to the other assumed
reconstructions.

Nevertheless, not all measurements agree with the ADI
model.\cite{Cha1999b, Tak2001, Kim2003} The main criticism to the ADI
model is due to the presence of a mirror plane, which should yield the
extinction of some diffraction reflections,\cite{Suz1999} because the
corresponding patterns have been seen in some
experiment.\cite{Myl1991} Thus, models different from the ADI one are
also being considered and a final conclusion has not been drawn yet.

An alternative approach for surface investigations is given by the
study of its optical properties. A very efficient technique is the
reflectance anisotropy spectroscopy (RAS), which can be used in situ
for the surface determination. The interpretation of such spectra is a
demanding task. For this purpose it is necessary to have theoretical
reference data together with a correlation of the spectral features
and the surface reconstruction elements.

The RAS of the Si(113) surface has been measured since ten
years.\cite{Man1996, Ros1996b} The first theoretical investigations of
the RAS for the Si(113) surface has been performed in 1998 by using a
dipole approximation for the ideal bulk atomic positions.\cite{Hog98}
In the same year, a tight-binding calculation has been done for the
Si(113)3$\times$2ADI model.\cite{Gav1998}, where the bulk
derivative-like peaks have been found with the wrong sign. Although
tight-binding calculations yield good results for bulk silicon, the
low-energy range of the spectra, which is related to the surface, is
usually not described very well.\cite{Pal99}

For this purpose, we choose an ab-initio approach to investigate the
optical properties of the Si(113) surface. The goal of this study is
to find out whether surface-related structures in the spectra can be
traced back to some characteristic surface reconstruction elements of
Si(113)3$\times$2ADI (pentagons, interstitial, adatoms, etc.), and
which kind of changes can be predicted with respect to other surface
models. In this way optical spectroscopy could be used to discriminate
between different models for reconstruction of the Si(113)
surface. Due to the wide acceptance of the Si(113)3$\times$2ADI, we
used this model as a starting point.

This article is organized as follows: After summarizing shortly the
methods used for the calculation (Section~\ref{Method}), we first
focus on the ground state properties (Section~\ref{GS}). Since the
optical spectra are determined by the electronic transitions, an
investigation of the electronic structure is presented in
Section~\ref{ES}. A detailed discussion of the optical spectra is
given in Section~\ref{Opt}, and, finally, we summarize and draw a
conclusion.

%
%
\section{Method}\label{Method}
We have performed ab-initio total energy calculations using the
periodic slab method within the framework of the density-functional
theory\cite{Hoh64} (DFT) as implemented in the ABINIT\cite{ABINIT} and
TOSCA\cite{TOSCA} packages. For the exchange-correlation energy in the
Kohn-Sham equations\cite{Koh65} the local-density
approximation\cite{Cep80,Per81} (LDA) has been chosen. The
eigenfunctions have been expanded into plane waves using
pseudopotentials, here normconserving ones in the Troullier-Martins
style.\cite{Tro91}

After converging the ground state structure with ABINIT, the optical
properties have been calculated using the TOSCA package.\cite{TOSCA}
The probability ${P}_{v{\bf k},c{\bf k}}^j$ of the electronic
transitions between the valence ($v$) and the conduction ($c$) states
with electronic eigenenergies $E_{c{\bf k}}$ and $E_{v{\bf k}}$ for
light polarized in $j$ ($j=x,y,z$) direction at a given point ${\bf
k}$ in the reciprocal space has been calculated as the diagonal
elements of the velocity operator.\cite{Bas75} Working within the
independent particle random phase approximation\cite{Ehr59} (IPRPA),
local-field, self-energy, and excitonic effects are neglected. For our
system we assume that they can be well described by the scissor
operator approach,\cite{Sol93} hence we stick to the IPRPA to describe
the optical properties. In order to obtain the RAS, in a first step
the imaginary part of the slab polarizability $\alpha^{\rm s}$ has
been calculated by
\\ \parbox{7.5cm}{\begin{eqnarray*}
  {\rm Im} [4 \pi \alpha^{\rm s}_{jj}(\omega)]
  &=&
  \frac{8 \pi^2 e^2}{m^2 \omega^2 A}
  \sum_{\bf k}
  \sum_{v,c} \left|{P}_{v{\bf k},c{\bf k}}^j\right|^2 \\
  &&
  \times \delta(E_{c{\bf k}}-E_{v{\bf k}}
  -\hbar \omega)
 \end{eqnarray*}}\hfill
\parbox{8mm}{ \begin{eqnarray} \label{EqImalpha}\end{eqnarray}} \\
for an energy $\omega$.\cite{Sol95} Here, $A$ is the surface area, $m$
and $e$ the electron mass and charge, respectively. In order to smooth
the resulting spectra, a Gaussian broadening has been applied, because
it leads to more well-defined structured spectra than the Lorentzian
one. The RAS is defined as the difference of the deviation from the
Fresnel reflectivity $\Delta R_j/R$ for the orthogonal polarizations
within the surface plane (here the $x$ and $y$ direction):
\begin{eqnarray}
{\rm RAS}= \frac{\Delta R}{R}=
 \frac{\Delta R_x}{R} -\frac{\Delta R_y}{R} \qquad .
\end{eqnarray}
The deviation from the Fresnel reflectivity for normal incident light
is given by\cite{Sol95}
\begin{eqnarray}\label{EqDrr}
  \frac{\Delta R_j}{R}
  = 4 \left(\frac{\omega}{c_0}\right)
  {\rm Im}\left[ 
  \frac{\alpha^{\rm surf}_{jj}(\omega)}{\alpha_{\rm b}(\omega)}
                              \right]
  \qquad ,
\end{eqnarray}
where $\alpha^{\rm surf}_{jj}$ is the complex surface polarizability
and $\alpha_{\rm b}(\omega)$ the bulk one ($c_0$: velocity of
light). The surface polarizability can be obtained by subtracting the
bulk contribution from the slab one. Since the imaginary part is taken
from the ratio, the bulk part in the numerator vanishes, and
Eq.~(\ref{EqDrr}) holds also for $\alpha^{\rm s}_{jj}$ instead of
$\alpha^{\rm surf}_{jj}$. The complex polarizability function $\alpha$
is derived from the imaginary part via the Kramers-Kronig transform.

Finally, to keep into account the self-energy and excitonic effects in
an approximative way, a scissor operator shift\cite{Sol93} has been
applied to the eigenenergies.
%
%
%
\section{Ground state}\label{GS}
For our calculations we have chosen the cell with the 3$\times$2ADI
surface reconstruction as described above. We have used slabs with 11
and 7 double layers (DL) of Si, respectively. The amount of vacuum was
taken equal to the slab thickness. Each DL consists of 12 Si atoms,
just the surface DL contains one atom less. Since the slab is not
symmetric, the dangling bonds at the bottom of the slab have been
saturated with hydrogen (H). Therefore, the cells contain 131 or 83 Si
atoms, respectively, and 18 H atoms. A top view of the slab is shown
in Fig.\ref{Pic_Structure}, while the side views of the 11~DL cell can
be seen in Figs.~\ref{Pic_chargeTamm} and \ref{Pic_LayerByLayer}.

The convergence of the total energy requires 4~{\bf k}
Monkhorst-Pack\cite{Mon76} points in the irreducible wedge of the
Brillouin zone (IBZ), and a kinetic-energy cutoff of 16~Ry. With these
parameters, the error in the total energy is less than 0.046~eV per
atom.

A relaxation of the topmost 6~DL, computing the forces acting on the
atoms (the remaining DL have been kept fixed to bulk positions) shows
significant changes with respect to the bulk positions just in the
topmost 4~DL. Therefore, these 4~DL have been optimized till the
forces are less than 0.08~eV/\AA.

We have calculated the surface energy by subtracting the energy of the
the hydrogenated surface and the bulk energy from the energy of the
slab. The resulting value of 12.74~eV per unit cell agrees very well
with the value of 12.69~eV obtained by Stekolnikov et
el.\cite{Ste2003a} As a consequence, the value of surface energy per
unit area of 87.87~meV/\AA$^2$ is in good agreement with the one of
Stekolnikov et al.\cite{Ste2003a} (87.36~meV/\AA$^2$). However, it is
slightly lower than the ones obtained from other groups
(97~meV/\AA$^2$[\onlinecite{Dab1994, Dab1995c}],
90.4~meV/\AA$^2$[\onlinecite{Lee2004}],
94.8~meV/\AA$^2$[\onlinecite{Lu2005}]) using various
methods. Nevertheless, it is lower than that for other surface
reconstructions and the overall agreement is good.

%
%
%
\section{Electronic structure}\label{ES} 
\begin{figure}[t]
  \epsfig{figure=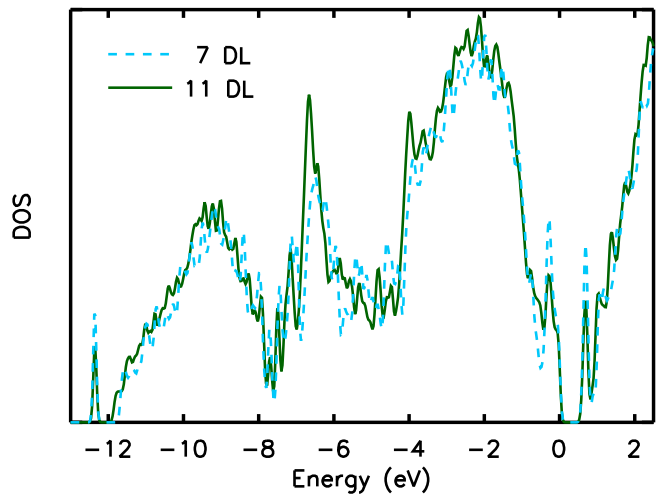,width=8.6cm}
  \caption{
    (Color online) Electronic density of states for the 11~DL cell
    (solid dark-green lines) and the 7~DL cell (dashed light-blue
    lines). The zero of the energy scale is set to the top-valence
    state, and the height of the DOS is scaled with respect to the
    total number of atoms in the corresponding cell (see text).
  }\label{Pic_DOS}
\end{figure}

After the ground state has been converged, a first step towards the
optical properties is an analysis of the electronic structure. We
started with the calculation of the electronic density of states (DOS)
as displayed in Fig.~\ref{Pic_DOS}, where the highest occupied band
(HOMO) is set to 0~eV. There, a comparison of the DOS calculated with
the 7~DL and the 11~DL slab is shown.  The DOS has been obtained using
a Brillouin-zone summation over 25~{\bf k} points, together with a
Gaussian broadening of 0.06~eV. The general shape is similar to the
one of bulk silicon for the occupied bands. In addition, around the
energy gap surface-related structures are visible: one strong peak at
the conduction band edge and a peak with a shoulder and another
shoulder close to the bulk-like part of the DOS at the valence band
edge. The use of the 7~DL and the 11~DL supercell leads to just minor
differences. By normalizing the DOS to the total number of atoms in
the slab, the height of the surface peaks (around the gap) appears to
be smaller for the thicker slab.

The valence peaks around the Fermi energy, which are in addition to
the bulk DOS, have been also found experimentally and they correspond
to surface bands. Using angle-resolved ultraviolet photoemission
spectra (ARUPS) Myler and Jacobi\cite{Myl1989} found two peaks
separated by 1.6~eV, where the second one is broad and could hide two
single peaks. A similar experiment has been performed using
angle-resolved photoelectron spectroscopy (ARPES) showing two single
peaks separated by 0.4~eV, and a third one at lower energy with a
non-dispersive character, which is not resolved in the other
experiment.\cite{An2001b, Hwa2001, Hwa2002} At normal emission, the
two close peaks fall together resulting in a single broad peak, which
has been found also by Myler and Jacobi.\cite{Myl1989} Thus, the
surface peak around the HOMO in the DOS is well defined in the
experiments, and it contains probably two peaks as indicated by the
shoulder. Furthermore, the core level spectra of Hwang et
al.\cite{Hwa2001, Hwa2002} were fitted using three surface peaks,
where the second shoulder in our DOS might refer to that third
peak. In scanning tunneling spectra (STS)\cite{Ara1998} two peaks are
visible. The low energy maximum has been attributed to the tetramers
and the high energy maximum to the adatoms of the Si(113)3$\times$2ADI
reconstructed surface. For the peak at the lowest unoccupied band
(LUMO),which should be accessible by inverse photoelectron
spectroscopy, no experimental results exist.

\begin{figure}[t]
 \epsfig{figure=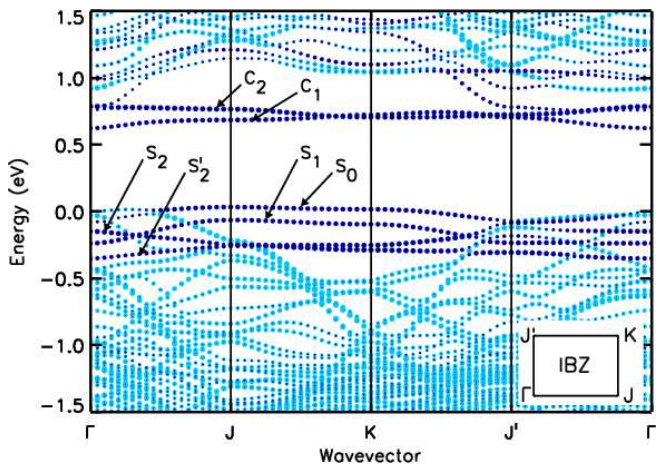,width=8.6cm,angle=0}
  \caption{(Color online) Dispersion of the electronic eigenenergies
  along the high-symmetry directions of the irreducible part of the
  Brillouin zone (IBZ). According to the
  localization of the corresponding states in the slab, light
  dots correspond to bulk states and dark ones to surface
  states. The dot size accords to the value obtained by
  Eq.~(\ref{Eq_loc}) (see text).
  }\label{Pic_Bands}
\end{figure}

After identifying the surface structures in the DOS we have
investigated the dispersion of the electronic eigenenergies using the
11~DL supercell. The results are displayed in Fig.~\ref{Pic_Bands}. In
order to separate the surface bands from the bulk background we have
calculated the localization $L_n$ for each state $n$ by
\begin{eqnarray}\label{Eq_loc}
  L_n = \int {\rm d}{\bf r} \psi^*_{n{\bf k}}({\bf r}) 
  \theta(z) \psi_{n{\bf k}}
\end{eqnarray}
where $\theta(z)$ is a realspace boxcar function\cite{Hog03}, meaning
$\theta(z)=1$ for a given range in $z$ direction and zero
otherwise. $\psi_{n{\bf k}}$ is the single band wave function for a
given point {\bf k} and the integral is taken over the whole
supercell. Considering the range, e.g., from the subsurface DL to the
vacuum above, the value of $L_n$ corresponds to the integrated charge
for that state located in the $x$-$y$ slice with the chosen $z$ range.
In this way, this value indicates the amount of localization in the
range selected by $\theta(z)$.

Fig.~\ref{Pic_Bands} shows the results. The surface-localized bands
are drawn with dark symbols (the $z$ range has been chosen to cover
the topmost two DLs), and the bulk-localized bands with light symbols
(the $z$ range has been chosen to cover the lowest seven DLs where the
relaxation has not changed the bulk positions). The size of the dots
corresponds to the amount of localization in this area. This means,
e.g., for the topmost four valence bands at the $\Gamma$ point, that
the top band is less localized in the bulk area than the second one
and the third band is less localized in the surface area than the
fourth one.

With this procedure we have identified three surface valence bands
(S$_0$, S$_1$, S$_2$) and two surface conduction bands (C$_1$, C$_2$),
as visible in Fig.~\ref{Pic_Bands}. All surface bands show a partial
overlap with the bulk band structure. This points to a reduced
influence of excitonic effects, and thus, the IPRPA scissor
approximation is expected to work well.  The state S$_2$ appears as a
state folded at J along the $[1\bar{1}0]$ direction, and hence shows
3$\times$1 periodicity. The other surface states show the 3$\times$2
one. Of course, the eigenvectors have 3$\times$2 symmetry for all
surface states due to the choice of the 3$\times$2 supercell. In both
the experiments, ARUPS and ARPES, valence surface states with
3$\times$1 and 3$\times$2 periodicity have been found, but with a
larger energy difference, which is here between J and K at about
0.16--0.19~eV. The shape of the bands S$_1$ and S$_2$ are in excellent
agreement with the measured ones, and show a similar dispersion of
about 0.17~eV and 0.20~eV, respectively, compared with the experiments
of An et al.\cite{An2001b} (S$_1$: 0.15~eV and S$_2$: 0.30~eV) and
Myler and Jacobi\cite{Myl1989} (S$_1$: 0.20~eV and S$_2$:
0.15~eV). Hence we conclude that these two states are the ones
appearing in the experiments. Along the $\Gamma$-J direction the
separation of the bands shows large variations within the two
experiments, where we have here an overlap. The discrepancy might be
due to the negligence of quasiparticle corrections as well as due to
experimental setups. Considering available theoretical investigations
of the dispersion of the eigenenergies of the Si(113) surface (besides
the tight-binding study of Wang et al.\cite{Wan1996} using a puckered
surface model, which is hence not comparable), there exist just the
investigation of Stekolnikov et al.\cite{Ste2003b}. Within their
results they obtain also the surface states S$_0$ and S$_1$, where the
S$_2$ might be hidden in the bulk-projected band structure. The energy
spacing between these states is similar to the one obtained here.

\begin{figure}[b] 
  \epsfig{figure=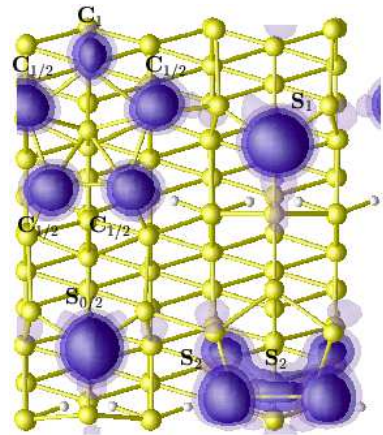,width=5cm,angle=0}
  \caption{(Color online): Top view ($x$--$y$ side like
    Fig.~\ref{Pic_Structure}) of the Si(113)3$\times$2ADI cell
    together with the charge density of the surface states S$_0$,
    S$_1$, S$_2$, C$_1$, and C$_2$ (see text). Note the overlap
    between the localization of some of the surfaces
    states.}\label{Pic_charge}
\end{figure}
By calculating the squared modulus of the wave functions of the surface
states we have been able to localize them more precisely in the slab
(see Fig.~\ref{Pic_charge}). The states S$_0$ and S$_1$ are located at
the adatoms (atom 3 and atom 9, respectively, see
Fig.~\ref{Pic_Structure}) and show a dangling bond character, which is
in agreement with the results of Ref.~[\onlinecite{Ste2003b}] and the
assumption of Ref.~[\onlinecite{Ara1998}]. The states C$_1$ and C$_2$
have been found at the atoms 4-7-8 and 5-6-7-8 of the pentagon with
the interstitial, which is stated similarly in
Ref.~[\onlinecite{Ste2003b}]. Furthermore, the surface state S$_2$ is
located at one adatom and two atoms of the empty pentagon (atoms 1-2
and 3), which has not been investigated before. However, this was
already assumed by Arabczyk et al.\cite{Ara1998}. Anyway, also this
state is present in experimental\cite{Kna1991, Dab1994, Dab1995c,
Sak1996, Zha2004c, Lee2004} and theoretical\cite{Dab1994, Dab1995c,
Ste2003b, Lee2004} STM images.

\begin{figure}[t] 
  \epsfig{figure=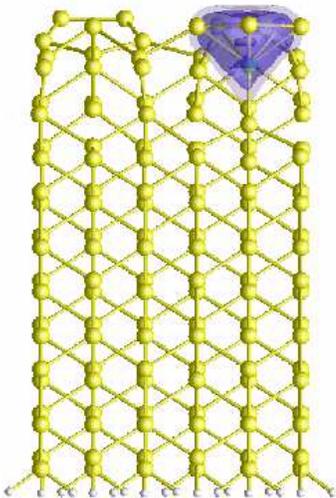,width=4.5cm,angle=180}
  \caption{(Color online): Side view ($x$--$z$ plane) of the
    Si(113)3$\times$2ADI cell together with the charge density of the
    low-energy Tamm-like state. Si atoms are drawn as large spheres, H
    atoms are as small spheres.
  }\label{Pic_chargeTamm}
\end{figure}
Besides the surface states above, we have obtained an additional
valence surface state at very low energy corresponding to the
separated peak at around $-12.5$~eV in the DOS of
Fig.~\ref{Pic_DOS}. This state shows a dispersion of less than 16~meV,
which follows the dispersion of the lowest valence states, which is 20
times larger. Thus, this nearly flat state can be considered as a
Tamm-like state.\cite{Tam32} This state is located in between the
atoms of the pentagon with interstitial (atoms 4-5-6-7-8 and 18), as
displayed in Fig.~\ref{Pic_chargeTamm}. The large separation between
these sites in the real space is responsible for the flat character of
the band.  However, it is not the first time that such a state has
been found. Pandey et al.\cite{Pan1975} found a flat surface band at
low energy ($\approx$~10.8~eV) which is partially below the bulk
valence bands for Si(111):SiH$_3$. Also a theoretical and experimental
investigation of the GaAs(110)1$\times$1 surface\cite{Pan1978} shows,
that one surface state is located below the valence band minimum with
a gap in between. The Tamm-like state found in our calculation should
be accessible within measurements and its experimental finding would
confirm the validity of the surface model used here. Due to its
low-energy character, transitions from that state will not appear in
the low or middle energy range of the RAS.
%
%
%
\section{Optical Properties}\label{Opt} 
In this section we present results for the optical spectra, in
particular for the RAS, of the Si(113) surface. We focus on the
convergence with respect to the {\bf k} points and the slab thickness,
we analyze the RAS regarding surface contribution and responsible
transitions, and we compare with available experimental results.
%
\subsection{Convergence tests}
For the optical properties we have performed convergence tests
concerning the number of conduction states and the number of {\bf k}
points used in Eq.~(\ref{EqImalpha}). 
\begin{figure}[b]
  \epsfig{figure=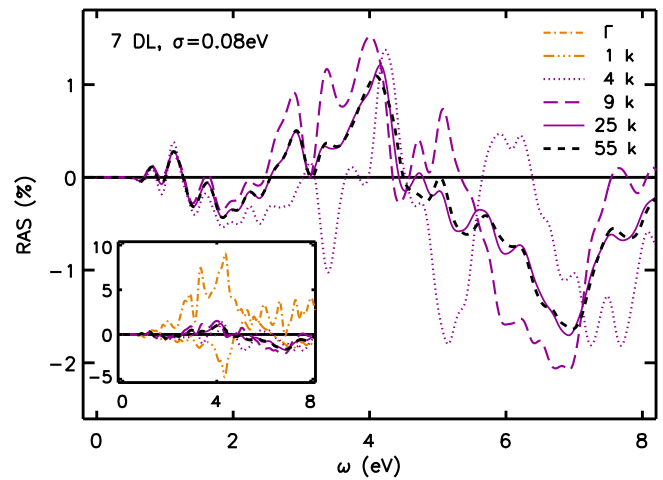,width=8.6cm}
  \caption{(Color online): Convergence of the RAS of
  Si(113)3$\times$2ADI with respect to the {\bf k} point
  summations. The 1 {\bf k} point sets ($\Gamma$ and Baldereschi) are
  shown in the inset, since the corresponding RAS is larger by one
  order of magnitude. Convergence requires 25~{\bf k} points in the
  IBZ.
  }\label{Pic_ConvK}
\end{figure}
A number of 460 bands in total (189 conduction states) have been
sufficient for the 11~DL cell as well as 360 bands (185 conduction
states) for the 7~DL one, respectively. Each state has been expanded
into plane waves. The use of 4000 {\bf G} vectors over the available
set of $\approx$24000 plane waves has been enough for the convergence
of the matrix elements. The convergence tests with respect to the
special {\bf k} points are more demanding. In our case, the {\bf k}
points can be chosen in one quarter of the BZ corresponding to the
IBZ.  A Gaussian broadening of 0.08~eV gives reasonably structured
spectra. We have performed tests with various sets of {\bf k} points
for the 7~DL cell as shown in Fig.~\ref{Pic_ConvK}.  As visible in the
figure, up to 25~{\bf k} points the spectra still show some
deviations. With a single {\bf k} analysis\cite{Gaa06c} it has been
figured out that the convergence in the $x$ direction is slower than
in $y$ direction and thus, an increase of the number of {\bf k} points
just in $x$ direction (resulting in the set of 55~{\bf k} points)
yields a strong improvement of the convergence. Comparing the spectra
based on 25 and 55~{\bf k} points we have concluded that convergence
has been already achieved with 25~{\bf k} points. Incidentally, using
the $\Gamma$ point only, a strong anisotropy appears (see inset of
Fig.~\ref{Pic_ConvK}) which can be traced back to infinite chains in
$x$ direction of the bulk layers.\cite{Gaa06c}.

\begin{figure}[t]
  \epsfig{figure=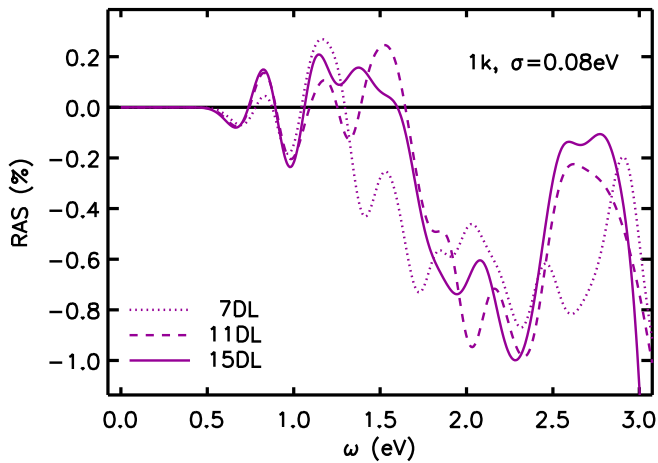,width=8.6cm}
  \caption{(Color online): Convergence of the RAS of
    Si(113)3$\times$2ADI with respect to the slab thickness (7~DL,
    11~DL, 15~DL) using only the Baldereschi point in
    Eq.~(\ref{EqImalpha}). Convergence is satisfactory using the 11~DL
    slab.
  }\label{Pic_ConvSlab}
\end{figure}

Since the slab thickness can influence the spectra, we have calculated
the RAS using only the Baldereschi\cite{Bal73} point for various cell
sizes, i.e. using slabs of 7~DL, 11~DL, and 15~DL. The 15~DL cell has
been created by adding four bulk DL to the 11~DL slab. The outcome of
the test is displayed in Fig.~\ref{Pic_ConvSlab}. It is clearly
recognizable that the 7~DL slab does not describe the low energy
region of the RAS correctly, where the differences between the 11~DL
and the 15~DL cell are just minor ones. Hence, the use of the 11~DL
cell produce reliable results for the surface spectra. Note that the
spectra using only the Baldereschi point are not converged, but the
use of just one point is sufficient for the check performed here. It
has been already shown that the convergence regarding the thickness
improves using more {\bf k} points.\cite{Sol99} Concerning the {\bf
k}-point convergence, we expect that our 25~{\bf k}-points set, which
has been shown to be enough for the 7~DL slab, will also be at
convergence for the 11~DL one.

%
\subsection{Analysis of the RAS}

We proceed now to analyze our fully converged RAS spectra (using the
11DL supercell and 25~{\bf k} points in the summation) with respect to
the surface structure. Thus, we have performed a layer-by-layer
decomposition as described in Refs.~[\onlinecite{Hog03, Cas03,
Mon03}]. For this purpose, a real space cutoff has been used in order
to separate the contributions coming from defined layers of the slab
by introducing a boxcar function $\theta(z)$ in the calculation of the
dipole matrixelements. The modified transition probability
$\tilde{P}_{v{\bf k},c{\bf k}}^j$ can be rewritten as
\begin{eqnarray}
  \tilde{P}_{v{\bf k},c{\bf k}}^j = 
  -i\hbar \int d{\rm \bf r} \psi^*_{v{\bf k}}
  \theta(z) \frac{\partial}{\partial r_j}
  \psi_{c{\bf k}}
\end{eqnarray}
and the summation of Eq.~(\ref{EqImalpha}) changes to
\\ \parbox{7.5cm}{\begin{eqnarray*}
  {\rm Im} [4 \pi \alpha^{\rm s}_{jj}(\omega)] &=&
  \frac{8 \pi^2 e^2}{m^2 \omega^2 A}
  \sum_{\bf k} \sum_{v,c} 
  \left[{P}_{v{\bf k},c{\bf k}}^j\right]^*
  \tilde{P}_{v{\bf k},c{\bf k}}^j\\
  && \times \delta(E_{c{\bf k}}-E_{v{\bf k}}
  -\hbar \omega) \qquad .
 \end{eqnarray*}}\hfill
\parbox{8mm}{ \begin{eqnarray} \end{eqnarray}} \\
Note that only one of P and $P^*$ must contain the $\theta$ function.
The decomposed spectra are presented in Fig.~\ref{Pic_CompExp}
together with a side view of the slab. We have numbered the DL from
the top to the bottom. For the analysis we have chosen three regions:
The topmost two DL (DL01 and DL02 (surface), while DL00 is the vacuum
layer above), the third and the fourth DL (subsurface, DL03 and DL04)
where the optimization changed the atomic positions with respect to
the bulk ones, and the bulk part together with the hydrogens (bulk,
DL05 till DL11, while DL12 is the vacuum layer below), since the
relaxation did not affect the atomic positions in DL05 and DL06.

\begin{figure}[b]
  \epsfig{figure=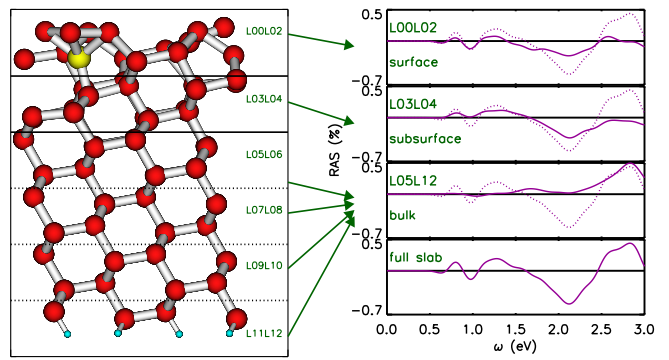,width=8.6cm}
  \caption{(Color online): Side view of the 11~DL slab (left side) and
    layer-by-layer analysis of the RAS (right side). Besides the
    decomposed spectra (solid lines) also the full spectra (dotted
    lines in the upper three panels and solid line in the lower one on
    the right ) is drawn for comparison. The arrows indicate where the
    contributions to the spectra come from. The DL assignment is
    explained in the text.
  }\label{Pic_LayerByLayer}
\end{figure}
In the low energy range of the spectra various peaks are visible. For
further discussion we have labeled the low energy peaks with P1, P2,
P3, P4, and P5 (see Fig.~\ref{Pic_CompExp}). Looking at the decomposed
spectra in Fig.~\ref{Pic_LayerByLayer} one can see that the the peaks
in the full spectra (containing the contribution of all DL) up to
2.5~eV (P1--P5) can be traced back to the surface and the subsurface
DL, where the structures at higher energies are also bulk
determined. Nevertheless, also at energies larger than 3~eV there
exist surface related structures. However, these structures are due to
surface-bulk, bulk-surface transitions and surface resonances. P1 and
P3 originate from the surface layers only, whereas the peaks P2, P4,
and P5 contain contributions from both the surface and the subsurface
layers. Thus we can conclude that for an analysis of P1 and P3 only
transitions between surface bands need to be considered, whereas for
P2, P4, and P5 also resonances or high-energy bulk states have to be
investigated.

After identifying the surface peaks in the spectra, we want to
identify the responsible transitions. Since for the deviation from the
Fresnel reflectivity Eq.~(\ref{EqDrr}) the whole complex
polarizability is considered, the RAS is also influenced by the real
part of $\alpha$. Thus, a overlap of spectral structures coming from
the imaginary and the real part is possible. A feasible procedure for
determining this overlap has been described in
Ref.~[\onlinecite{Hog05a}], where Eq.~(\ref{EqDrr}) was rewritten as
\begin{eqnarray}\label{EqCoefficients}
  \frac{\Delta R_j}{R}= A \, {\rm Im}[4\pi\alpha_{jj}^{\rm s}]
  +B\,  {\rm Re}[4\pi\alpha_{jj}^{\rm s}] \quad .
\end{eqnarray}
The energy-dependent coefficients $A$ and $B$ are determined only by
the dielectric function of the bulk crystal. With the magnitude of $A$
and $B$ one can figure out if the RAS is mainly determined by ${\rm
Im}(4\pi\alpha^{\rm s})$ or by ${\rm Re}(4\pi\alpha^{\rm s})$. In this
way, $A$ and $B$ are surface independent coefficients.

\begin{figure}[t]
    \epsfig{figure=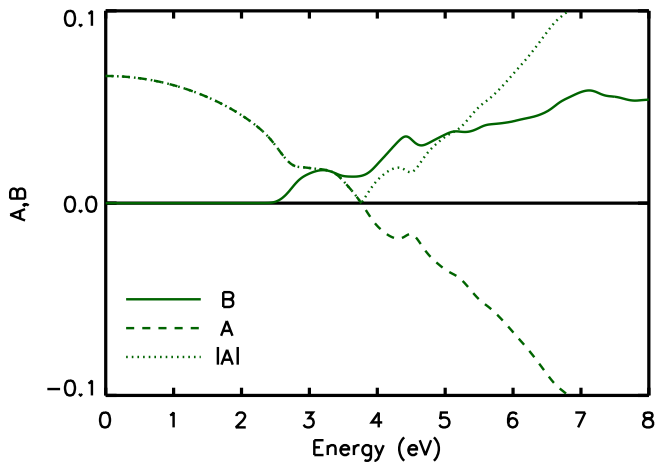,width=8.6cm}
    \caption{(Color online): bulk coefficients A and B from
    Eq.~(\ref{EqCoefficients}). For energies smaller than 2.5~eV there
    is no contribution of ${\rm Re}(4\pi\alpha^{\rm s})$ to the
    RAS.
    }\label{Pic_BulkCoefficients}
  \end{figure}

Because of this, we analyzed these coefficients for our system, in
detail, we determined $A$ and $B$ for bulk silicon. They are universal
for all silicon surfaces, independent of the orientation and
reconstruction. The result is shown in
Fig.~\ref{Pic_BulkCoefficients}. Besides $A$ and $B$, also $|A|$ is
plotted in order to compare the magnitude of the coefficients.
The result is the following: Every RAS of any clean silicon surface is
determined only by ${\rm Im}(4\pi\alpha^{\rm s})$ for an energy up to
2.5~eV and by both, ${\rm Im}(4\pi\alpha^{\rm s})$ and ${\rm
Re}(4\pi\alpha^{\rm s})$, for higher energies. In detail, it is mainly
determined by ${\rm Im}(4\pi\alpha^{\rm s})$ for 2.5--3.2~eV, by both
equally from 3.2--3.4~eV, mainly by ${\rm Re}(4\pi\alpha^{\rm s})$ for
3.4--5.1~eV, and by $-{\rm Im}(4\pi\alpha^{\rm s})$ for higher
energies. Thus, the surface peaks P1--P5 can be traced back to the
differences just between ${\rm Im}(4\pi\alpha_{xx}^{\rm s})$ and ${\rm
Im}(4\pi\alpha_{yy}^{\rm s})$. These are the differences we
investigate now in order to find the origin of the surface features in
the RAS.

Since the RAS in the low-energy range (up to 1.3~eV, peaks P1 to P3)
has been reproduced already sufficiently well using the Baldereschi
point only, based on this spectra the contribution of various
transitions to the first three surface peaks has been analyzed. For
the other spectral structures all {\bf k} points have been taken into
account.

The first peak P1 of the RAS displayed in Fig.~\ref{Pic_CompExp}
appears due to the transition from the surface valence state S$_0$ to
the surface conduction state C$_1$. Considering the localization of
the states (see Fig.~\ref{Pic_charge}), namely the position of the
adatom 3, where the charge density of S$_0$ has been found, and that
of the atoms 4-7-8 (the location of C$_1$) one sees that they are much
nearer in $y$ direction than in $x$. Therefore the transition
probability in this case is higher for $y$ polarized light, resulting
in a negative RAS signal. Note that the transition from S$_0$ to C$_2$
is forbidden.  Also the peak P2 in the RAS is due to transitions
between surface bands: the transitions from S$_1$ to both the surface
conduction bands C$_1$ and C$_2$ are responsible for it. Here, the
distance between adatom 9 (S$_1$) and the filled pentagon 4-5-6-7-8
(C$_1$ and C$_2$) in $x$ direction is small, where also a little
overlap of the charge density appear. This yields in a strong positive
signal in the RAS. Note that due to the overlap of S$_1$ with the
bulk bands there is also a tiny contribution of bulk-surface
transitions to P2.  The peak P3 is determined by transitions from
S$_2$ to C$_1$ and C$_2$. The vicinity of the state S$_2$, which is
located at atoms 1-2-3, to the states C$_1$ and C$_2$ at the pentagon
with interstitial (4-5-6-7-8) is mainly due to its localization at
adatom 3. From the charge density localized at the atoms 1-2, for both
polarizations the transition probability should be similar since the
distances in $x$ and $y$ directions are equal. The additional
localization at adatom 3 favor the $y$ direction yielding the negative
RAS signal. For the remaining two peaks P4 and P5 a mixture of
bulk-surface and surface-bulk transitions (not restricted to single
surface bands) is responsible, therefore a clear picture could not be
drawn.  The peaks P2 and P3 are equal in height, whereas P1 is a lower
one. This might be due to the fact that for P2 and P3 transitions to
two surface conduction states are responsible, but one of them is
forbidden for P1. However, the overall magnitude of the probability of
single transitions between surface bands is nearly the same in this
case.

Considering the dispersion of the eigenenergies, the regions in the
IBZ with the main contributions to the RAS have been identified using
again a single-{\bf k} analysis\cite{Gaa06c}: For P1, {\bf k} points
near $\Gamma$ carry the main contribution, whereas for P2 {\bf k}
points close to the J-K boundary (large $k_x$ component), and for P3
those close to the $\Gamma$-J boundary (small $k_y$ component)
comprise the main contributions. Such a trend has not been found
neither for P4, nor for P5. A posteriori, due to this anisotropic
distribution of the contribution, the convergence with respect to the
{\bf k} points had to be expected to be a delicate one.

Changing the surface reconstruction certainly would affect the RAS. We
can speculate that removing, e.g., the interstitial atom (atom 18)
would change the localization of the states C$_1$ and C$_2$ and which
will have an effect to all three peaks discussed here. Adding an
interstitial to the empty pentagon would affect mainly state S$_2$,
meaning the third peak in the spectra. Furthermore, having two nearly
equivalent pentagons on the surface would influence the peak height,
because the transition probability for the $x$ and $y$ would not be as
significantly different as for the ADI reconstruction. However, also
the energetic position of the surface peaks would indeed be affected.

%
\subsection{Comparison with experimental results}
\begin{figure}[b]
  \epsfig{figure=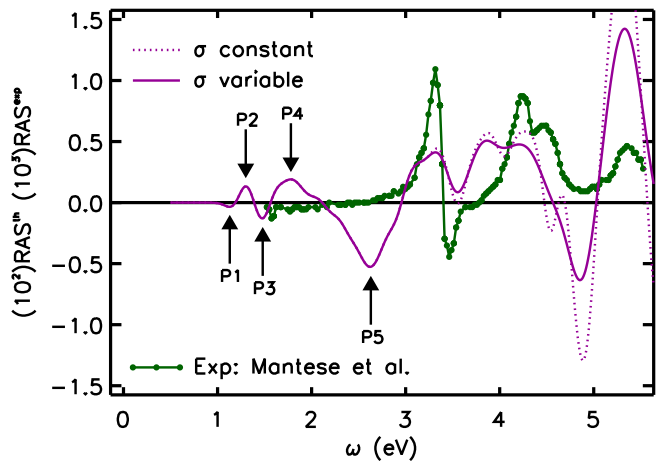,width=8.6cm}
  \caption{(Color online): Comparison of the RAS calculated using
  25~{\bf k} points in the 11~DL cell (light curves) obtained with a
  constant broadening of 0.08~eV (dotted line) and a variable one
  (solid line), see text. The experimental spectra are from
  Refs.~[\onlinecite{Man1996, Ros1996b}] (dark dots connected by a
  solid line). A scissor operator of 0.5~eV has been applied to the
  theoretical spectra, and the RAS amplitude has been scaled by a
  factor of 10 (see text).
  }\label{Pic_CompExp}
\end{figure}
Finally, in Fig.~\ref{Pic_CompExp} we compare our RAS with the
experimental data obtained by Aspnes and coworkers\cite{Man1996,
Ros1996b}. Unfortunately, the measured RAS did not show any
surface-related spectral structure. This might be due to passivation
of the surface with Si-O suboxide. Thus, the theoretical surface peaks
we have investigated cannot be compared. Regarding bulk-related
structures, using a 11~DL slab does not allow the perfect description
of the bulk derivative-like structure. In order to wash out artificial
oscillations of bulk structures, we have applied a variable
broadening. This means, we have employed a constant broadening for
energies smaller than 2.5~eV (inset of the bulk spectra) and an
additional broadening which goes linearly with the energy (factor
0.04) for higher energies.  Furthermore, we scaled the measured RAS
with a factor of 10, which is the usually accepted factor for silicon
surfaces.\cite{Inc2006} In this way we got a reasonable agreement with
respect to our slab thickness in the medium energy range. As visible,
the calculation reproduces the spectral structure at around 3.2~eV and
the double peak at 4.2~eV, which is shifted slightly to lower energies
in our case. Also the experimental structures at around 5~eV can be
found in our spectra, however, it is a little bit overestimated. The
bulk derivative-like structure itself at around 3.5~eV is reproduced
only qualitatively, which is a result of the small slab size. Compared
to the tight-binding calculation of Gavrilenko et al.\cite{Gav1998},
our spectra can resolve the characteristic low-energy surface peaks
and describes the bulk derivative-like structure with the correct
sign.

%
%
%
\section{Summary and Conclusions}\label{Summary}

In summary, we have investigated the electronic structure and the RAS
spectrum of the Si(113)3$\times$2ADI. The electronic band structure
and the DOS were found in excellent agreement with available
experimental results.  We identified three surface valence bands and
two surface conduction bands near the Fermi level by computing the
wave function localization in the surface region. The valence surface
states are located at the adatoms and at the pentagon without an
interstitial, whereas the conduction surface states are located at the
pentagon with an interstitial, in agreement with STM images. The three
valence surface states in the energy gap compare with the two states
which are visible in ARUPS\cite{Myl1989} and the ARPES\cite{An2001b,
Hwa2001, Hwa2002} experiments, where in the latter one a third surface
(resonance) state has been found in overlap with the bulk
states. Nevertheless, the existence of three surface valence states
has been confirmed by the comparison of theoretical\cite{Dab1994,
Dab1995c, Ste2003b, Lee2004} and experimental\cite{Kna1991, Dab1994,
Dab1995c, Sak1996, Zha2004c, Lee2004} STM images and by a core-level
analysis.\cite{Hwa2001, Hwa2002} Two of the valence surface states
show a 3$\times$2 and one a 3$\times$1 periodicity, which has been
confirmed experimentally.\cite{Myl1989,An2001b} In addition, we found
a surface state below the valence band minimum. This state is located
at the filled pentagon and shows a Tamm-like character. This finding
calls for more experimental work for measuring the DOS at low energy.

Concerning optical spectra, after checking the numerical convergence
with respect to the slab thickness and the number of {\bf k} points,
we have performed a layer-by-layer spectral decomposition. In this way
the surface-relevant spectral features have been
determined. Furthermore, an analysis of the contributions of the real
and the imaginary part of the polarizability lead to the conclusion
that the surface relevant structures are due to the imaginary part
only. We have traced back the first three low-energy peaks of the RAS
to transitions from the three valence surface states S$_0$, S$_1$, and
S$_2$ (localized at the adatoms and the empty pentagon) to both the
conduction surface states, C$_1$ and C$_2$ (localized at the pentagon
with interstitial). These RAS peaks are hence very sensitive to a
change of the surface reconstruction. The missing splitting of the
peaks in the RAS with respect to the two surface conduction states is
assumed to be due to the small energy difference between them. The
only available experimental RAS does not show any low-energy
surface-related spectral features, probably due to a surface
contamination. At higher energy, the bulk-derivative like structure,
as well as other spectral features, agree with those of the measured
RAS. Differences with respect to the experimental results can be due
to the limited size of the slab used in the calculations and to the
negligence of self-energy and excitonic effects.

%
%
%
\section*{Acknowledgment}
This work was funded by the EU's 6th Framework Programme through the
NANOQUANTA Network of Excellence (NMP4-CT-2004-500198).  We would like
to thank A.~Incze, C.~Hogan, and R.~Del~Sole for fruitful
discussions. K.~GN wants also to thank S.~Hinrich for providing
articles and additional information about the Si(113)
surface. Computer facilities at CINECA granted by INFM (Project
no. 352/2004 and no. 426/2005) are gratefully acknowledged.
%
%
%
\bibliography{Si113,Theory,Opt,MyAndOthers}

\end{document}